\documentclass[10pt]{iopart}
\input epsf
\begin{document}
\jl{3}
\title[Free Energy Landscape Near The Glass Transition]
{Free Energy Landscape Of Simple Liquids Near The Glass Transition}
\author{Chandan Dasgupta\dag \footnote[3]{Also at the Condensed 
Matter Theory Unit, Jawaharlal 
Nehru Centre for Advanced Scientific Research, Bangalore 560064,
India.} and Oriol T. Valls\ddag}
\address{\dag\ Centre for Condensed Matter Theory, 
Department of Physics, Indian Institute of Science, Bangalore 
560012, India}
\address{\ddag\ School of Physics and Astronomy and Minnesota 
Supercomputer Institute, \\ University of Minnesota, 
Minneapolis, Minnesota 55455, USA}
\begin{abstract}
Properties of the free energy landscape in
phase space of a dense hard sphere system characterized by a discretized
free energy functional of the Ramakrishnan-Yussouff form are
investigated numerically. A considerable number of glassy local minima 
of the free energy are located and  the distribution of
an appropriately defined ``overlap'' between  minima is calculated.
The process of transition  from the basin of attraction of 
a minimum to that of another one is studied using a new
``microcanonical'' Monte Carlo procedure, leading to a determination
of the effective height of free energy
barriers that separate different glassy minima. 
The general appearance of the free energy landscape resembles 
that of a putting green: deep minima separated by
a fairly flat structure. The growth of the effective
free-energy barriers with increasing density is consistent with 
the Vogel-Fulcher law, and this growth is primarily driven by an
entropic mechanism.
\end{abstract}
\submitted{{\noindent \it }}

\section{Introduction}
\label{sec:intro}
A liquid quickly cooled to temperatures below its  freezing
point  enters a metastable supercooled state. At lower 
temperatures, the supercooled liquid undergoes a
glass transition to a state in which it resembles
a disordered solid. The dynamics of supercooled liquids near
the glass transition exhibits~\cite{rev1,rev2} multi-stage, non-exponential 
decay of fluctuations and a rapid growth of relaxation times,
features which are not fully understood.

An intuitive description that
is often used~\cite{pwa,pgw} for  a qualitative understanding of the
observed behavior near the glass transition is based on the
``free energy landscape'' paradigm. 
This description starts from a functional that expresses the free
energy of a liquid in terms of the time-averaged local number
density. At high temperatures (or at low densities in systems, such as
hard spheres, where the
density is the control parameter), this  functional is
believed to have only one minimum, that representing the uniform liquid
state. As the temperature is decreased to near the 
crystallization point, a new minimum representing the crystal,
with a periodic modulation of the local density, 
should also develop. In the ``free energy landscape'' picture, 
a large number of ``glassy'' local minima
of the free energy, characterized by inhomogeneous, aperiodic
density distributions, also appear at temperatures below
the equilibrium freezing point. If the system gets 
trapped in one of these glassy local minima as it is cooled
rapidly, crystallization can not occur and
the subsequent dynamics is governed by
thermally activated transitions among some of the many 
metastable glassy minima. If the system visits many of these minima
during its evolution over a certain observation time, 
it behaves like a liquid
over such time scales: the time-averaged local density
remains uniform. However, the dynamics in this regime, 
governed by thermally activated transitions, is 
slow and complex. In this picture,
the glass transition occurs when the time scale of transitions among 
the glassy minima becomes so long that the system is confined in
a single ``valley'' of the landscape over experimentally accessible 
time scales. The features of such a free energy landscape would be
very similar to those found~\cite{kw,kt,ktw} in
certain generalized spin glass models with infinite-range interactions,
and in spin models~\cite{bm,parisi} with 
complicated infinite-range interactions, but no quenched
disorder. The behavior of these mean-field models exhibits a remarkable
similarity with the phenomenology of the glass transition.
These results suggest that the free energy landscape paradigm 
indeed provides a good framework for the 
understanding of the properties of supercooled liquids near the glass
transition. Such a description requires numerically
obtained  information about the topography of the free energy
landscape of these liquids. 

We have carried out several numerical studies of a dense hard-sphere 
system, using a model free
energy functional proposed by Ramakrishnan and Yussouff (RY)~\cite{RY},
a discretized version of which exhibits~\cite{cd1,cd2}
a large number of glassy local minima at densities
higher than the value at which equilibrium crystallization 
occurs. [The control parameter for a hard-sphere system is the dimensionless 
density $n^* \equiv \rho_0 \sigma^3$, where $\rho_0$ is the average
number density in the fluid phase and $\sigma$ is the hard-sphere
diameter; increasing 
(decreasing) $n^*$ has the same effect as decreasing (increasing) 
the temperature where the temperature is the 
control parameter.] From numerical studies~\cite{lvd,dv94,vd95}
of the Langevin equations for this system, we found that the 
dynamics changes qualitatively 
at a ``crossover'' density near $n^*_x = 0.95$. The dynamics of a
system initially  in the uniform liquid state remains
governed by small fluctuations near the  liquid free energy minimum 
when the density is lower than $n^*_x$. 
For higher values of $n^*$, the dynamics
is governed by transitions among the glassy minima. 
The time scales for such transitions were estimated 
from a  Monte Carlo (MC) method in Ref.~\cite{dv96} and found to 
increase rapidly with density. 

Here we report results of additional numerical studies in
which a new approach to the free energy landscape is used. 
We have developed and
used a new MC procedure that enables us
to study transitions between different glassy
minima and thus investigate
the topography of the free energy surface in phase space. We have 
located a large number of glassy minima of the free energy so
as to yield their statistical properties as a function
of density. The total number of glassy minima is found to remain 
nearly constant as the density
is varied in the range $0.94 \le n^* \le 1.06$. The free energies of the
glassy minima are distributed over a wide range between the free
energy of the uniform liquid and that of the crystal. An appropriately
defined ``overlap'' between different glassy minima is also found to
exhibit a broad distribution. We have found pairs of glassy minima that differ
from each other in the rearrangement of a very small number of
particles. The height of the free energy barrier that separates two
such minima is quite small. Such pairs
may be identified as ``two-level systems'' which are believed~\cite{tls}
to exist in all glassy systems. 

Our computation of the probability of transition from a glassy minimum
to the others as a function of the free energy increment (see below) 
and the MC
``time'' $t$ leads us to define an effective barrier height that
depends weakly on $t$. The growth of this 
effective barrier height with  density is consistent with a 
Vogel-Fulcher form~\cite{vf} for a hard-sphere system~\cite{wa81}. 
The dependence of the effective barrier height on $t$ and the
density indicates that the growth of 
the barrier height (and the consequent growth of the relaxation time)
is primarily due to entropic effects arising from an increase in the 
difficulty of finding low free-energy paths (saddle points) that 
connect one glassy local minimum with the others. 

\section{Model and Methods}
\label{section:m&m}

We characterize our system 
by a free energy functional $F[\rho]$  of the
form~\cite{RY}:

\begin{eqnarray}
F[\rho] &=& F_l(\rho_0)+ k_B T \left[ \int{d {\bf r}\{\rho({\bf r})
\ln (\rho({\bf r})/\rho_0)-\delta\rho({\bf r})\} } \right. \nonumber \\
&-& \left. (1/2)\int{d {\bf r} \int {d{\bf r}^\prime
C({|\bf r}-{\bf r^\prime|}) \delta \rho ({\bf r}) \delta
\rho({\bf r}^\prime)}} \right],
\label{ryfe}
\end{eqnarray}
where $F_l(\rho_0)$ is the free energy of the uniform liquid at density
$\rho_0$, and $\delta \rho ({\bf r})\equiv \rho({\bf r})-\rho_0$ is the
deviation of the density $\rho$ at point ${\bf r}$ from $\rho_0$.
We set $F_l(\rho_0)=0$. In Eq.(\ref{ryfe}),
$T$ is the temperature and $C(r)$ the direct pair correlation
function~\cite{hm86} of the uniform liquid at density $\rho_0$, which
we express in terms of 
$n^*\equiv \rho_0 \sigma^3$ ($\sigma$ is the hard-sphere diameter) 
by making use of the Percus-Yevick~\cite{hm86} approximation. 
The direct pair correlation function~\cite{hm86}  of simple model
liquids characterized by an isotropic, short-range pair-potential with 
a strongly repulsive core (such as the Lennard-Jones potential) is very
similar to that of the hard-sphere system at high densities. Therefore,
we expect our results to apply, at least
qualitatively, to such dense liquids.

We discretize our system
by introducing a cubic lattice of size $L^3$ and mesh constant
$h$ in which the variables 
$\rho_i,\, i= 1, L^3$, are defined as
$\rho_i \equiv \rho({\bf r}_i) h^3$, where $\rho({\bf r}_i)$ is the
density at mesh point $i$.
The dimensionless free energy per particle $f[\rho]$ is
$f[\rho]= \beta F[\rho]/N$
where $N = \rho_0 (Lh)^3 = n^* L^3 a^3$ is the total number of
particles in the simulation box, $\beta \equiv 1/(k_BT)$ and 
$a$ is the ratio $h/\sigma$.

Ideally, one would like to start
the system in a known  glassy local minimum
of the free energy, and 
investigate the topography of the free energy surface near the starting
point by allowing the system to evolve, and finding out which
configurations it subsequently visits and where it ends up. 
A conventional Metropolis algorithm MC
procedure~\cite{dv96} is inefficient at doing 
this because at the relatively high
densities studied here, it would take a very long time for the
system to move out of the basin of attraction of the initial minimum.
To obviate this difficulty, 
we have devised what we call a ``microcanonical'' MC
method. The algorithm is as
follows: we choose a trial value of what we call the free energy increment,
$\Delta F$, or $\Delta f$ if we are 
dealing with the dimensionless version of $F$. Then, starting with
initial conditions which correspond to a
local free energy minimum, we sweep the sites $i$ of the lattice sequentially.
At each step and site, we pick another site $j$ at random from the ones
within a distance $\sigma$ from $i$. 
We then attempt to change the values
of $\rho_i$ and $\rho_j$ to $p(\rho_i+\rho_j)$ and $(1-p)(\rho_i+\rho_j)$,
where $p$ is a random number distributed uniformly in $[0,1]$. 
The attempted change
is accepted, and this is the crucial point, if and only if the free energy
after the change is less than $F_{max}\equiv F_0 +\Delta F$ where
$F_0$ is the  value of the free energy at
the minimum where we start the computation.
The simulation proceeds up to a maximum ``time'', $t_{m}$,
measured in MC steps per site (MCS). We
perform a sweep over a range of values of $\Delta F$, with the same
initial conditions. If $\Delta F$ is smaller than the height of 
the lowest free energy barrier between the starting minimum
and any other ``nearby'' minima, the system will
remain in the basin of attraction of the starting minimum. As we
increase $\Delta F$, there will eventually be 
one or more minima that the system can find within 
a ``time'' $t<t_m$. These minima
are separated from the initial minimum by free energy barriers of
height less than $\Delta F$.
As $\Delta F$ is further increased, additional minima
will be made accessible, and since additional paths will become
available between the initial minimum and the minima already accessible at
smaller values of $\Delta F$, these minima may be reached in fewer MC steps.
Clearly, if one obtains the information of near which minimum the
system is, and how long it takes to get there, one can begin to map out
the free energy landscape.

To find out which basin of attraction the system is in at
time $t$, we save the  values of the 
variables $\rho_i$ at  relatively
frequent time intervals $\Delta t$. These configurations are then used
as the inputs in a minimization procedure~\cite{cd1} that determines which
basin of attraction the system is in. 
The entire procedure is repeated
a  number of times (the ``number of runs'') and averaged over.
We have carried out this  procedure  at 
densities in the range $0.94 \le n^* \le 1.06$. We did not consider
densities lower than 0.94 because previous studies~\cite{dv94,vd95} 
show that the dynamics of the system is governed by transitions among
glassy local minima only at higher densities. Since
the Percus-Yevick approximation  becomes 
less accurate at relatively high densities~\cite{hm86}, 
values of $n^* > 1.06$ were not considered.

We used two different sets of the sample size $L$ and 
the mesh size $h$, in one case
commensurate with a close-packed lattice, and in the other incommensurate.
The computationally more intensive part of our 
simulations was carried out for systems of size $L = 15$ 
with  periodic boundary conditions and  mesh size $h = \sigma/4.6$. 
No crystalline
minimum was found for these incommensurate values. 
The other portion of the computations was performed for systems
with  $L = 12$
and $h = 0.25 \sigma$. These values are commensurate with a
fcc structure and a crystalline minimum is found at
sufficiently high densities. Because of the smaller size of these
samples, we were
able to explore more extensively several
aspects of the problem under consideration.
Our computations for the $L=15$ sample were carried out for 
$t_m=15000$ MCS and $\Delta t=5000$ MCS, whereas computations for 
the system with $L=12$ were carried out to $t_m=8000$ MCS with 
$\Delta t=2000$ MCS. A detailed discussion of how the glassy minima
used in our study were chosen and the structure of these glassy minima
may be found in Refs.~\cite{dv96,dv98,dv99}.

\section{Results}
\label{section:r&d}

During the evolution
of the system, we monitor $\beta F$ and
the maximum and minimum values
of the variables $\rho_i, i = 1,L^3$.  If the system
fluctuates near one of the inhomogeneous minima,
then the maximum value of $\rho_i$ would be much higher than the value
(close to $\rho_0 h^3$) it would have in the vicinity
of the uniform liquid minimum. The system does not move to
the neighborhood of the liquid minimum for the values of $\Delta F$
considered here. The total free energy remains nearly
constant at a value slightly lower than the maximum allowed value, 
$F_{max} = F_0 + \Delta F$. 

In our analysis of the process of transition of the system 
from the initial glassy minimum to the basins
of other minima, we define a ``critical'' value, $\Delta f_c(t)$ (or
$\Delta F_c(t)$), of the free-energy increment $\Delta f$ (or $\Delta
F$) as follows:
at every time investigated (i.e. times 5000, 10000 and 15000 MCS 
for $L=15$
and times 2000, 4000, 6000 and 8000 MCS for the $L=12$ samples),
we test, for increasing
values of $\Delta f$, what is the probability,
$P(\Delta f,t)$, that the
system has moved to the basin of attraction of a free energy minimum
distinct from the starting one. This probability, which 
we obtain by averaging over a sufficient number (ten
to fifteen) of runs, is (at constant time)  zero
for very small $\Delta f$ and rises toward unity as $\Delta f$ 
increases. At a constant $\Delta f$, it increases somewhat with 
MC time, as the system  explores further regions
of phase space. We define $\Delta f_c(t)$ as the value of $\Delta 
f$ for which, at that time, the switching probability reaches $1/2$. 
Of course, $P$ and $\Delta f_c$ are also functions
of $n^*$. The minima to which the system moves for values of 
$\Delta f$ close to
or higher than $\Delta f_c$ are, in general, different for
different runs. This suggests that $\Delta f_c$ represents
a measure of the free energy increment for which a 
relatively large region of
phase space becomes accessible to the system. 
The system almost never returns to
the basin of attraction of the initial minimum: after having
left the initial minimum, the system cannot find
its way back. 

\begin{figure}[htbp]
  \begin{center}
    \leavevmode 
    \hspace*{1.8cm}
    \epsfysize=5.5cm\epsfbox{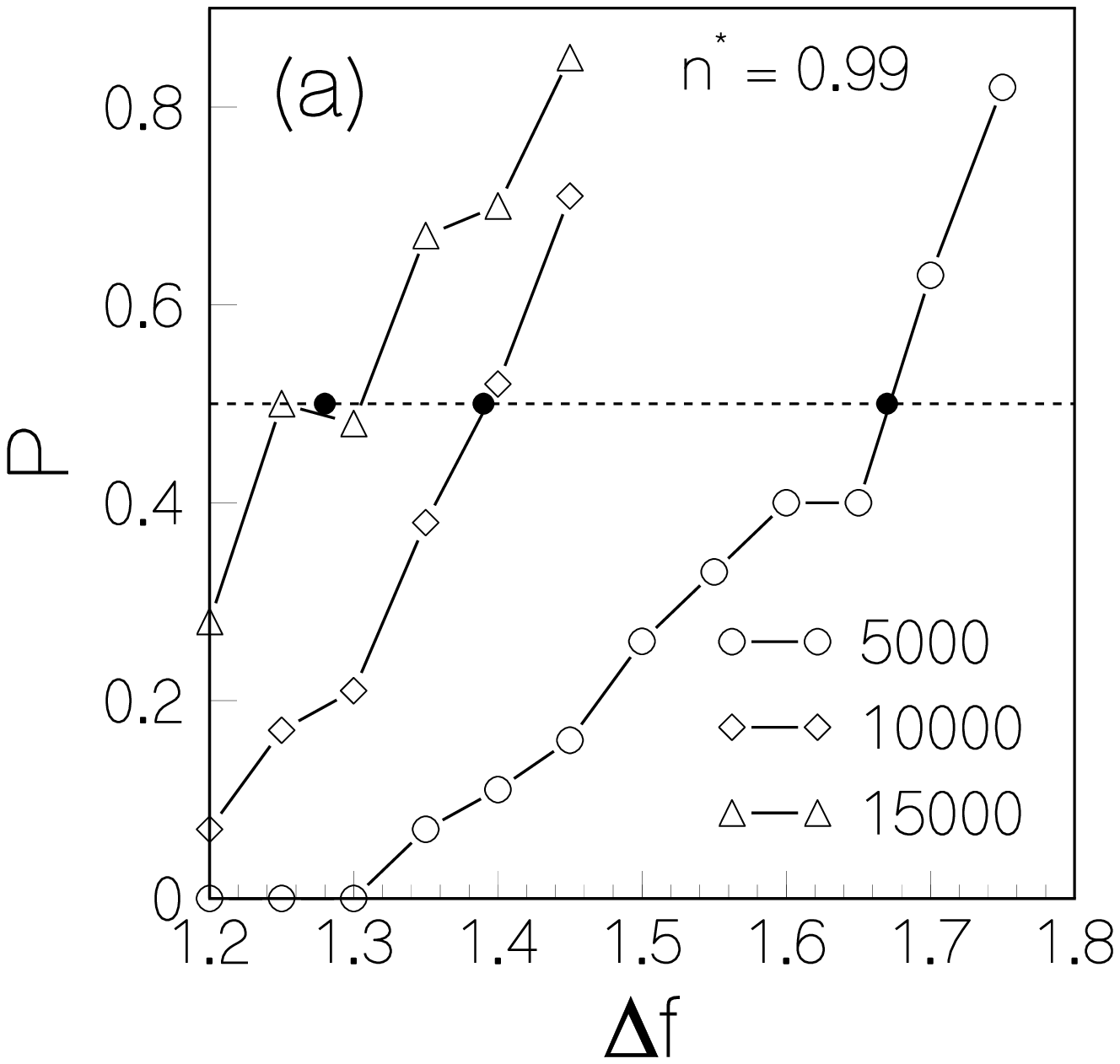}
    \epsfysize=5.5cm\epsfbox{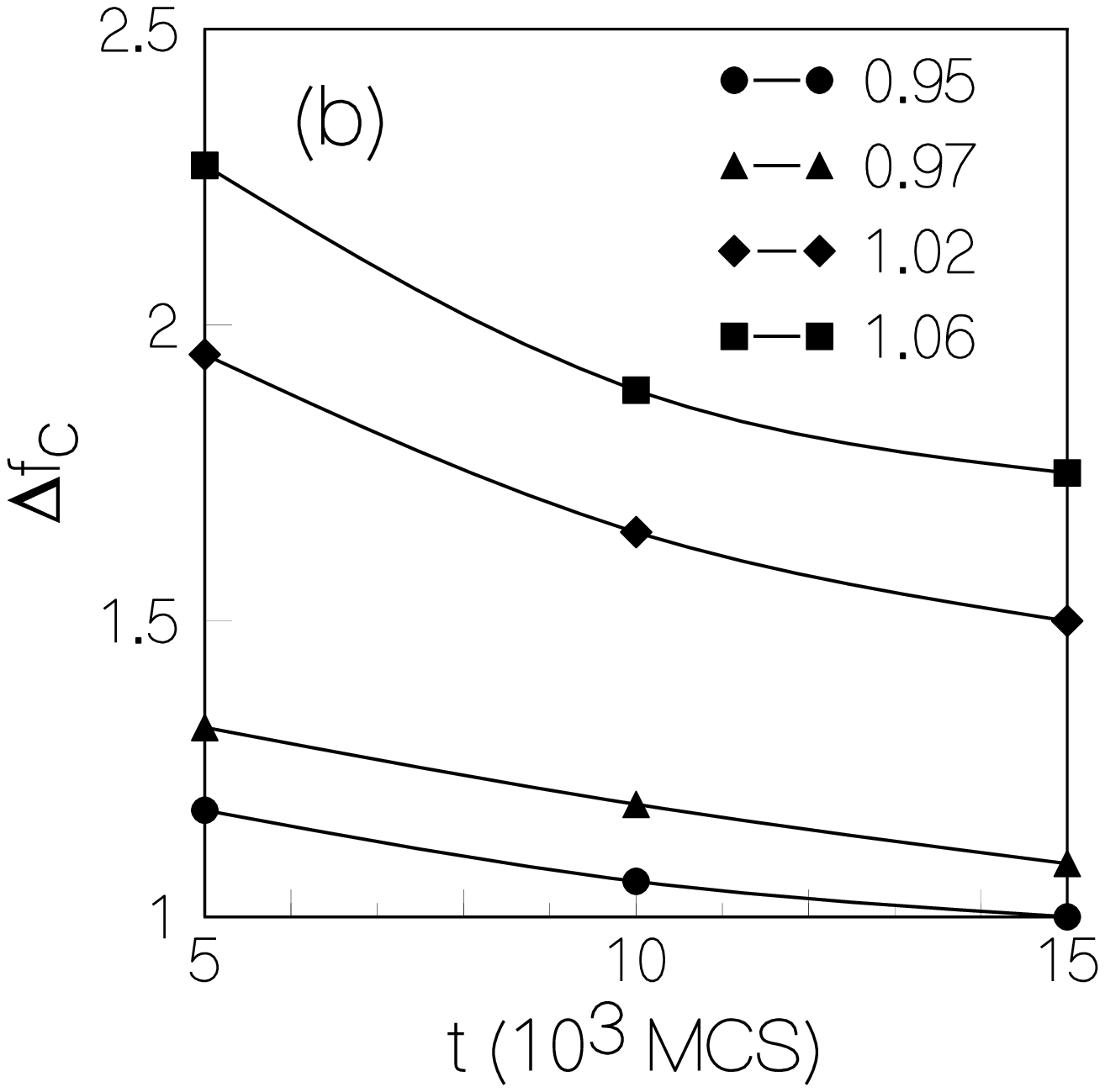}
  \end{center}
  \caption{ (a) Example of the determination of the 
``critical'' value $\Delta f_c$, defined as the value of the free
energy increment $\Delta f$ at which the transition probability $P$ is 1/2.
The black dots mark the intersections of the plots with
the line $P=0.5$ (see text for a complete discussion). The data shown
are for a sample of size $L = 15$, and three values (5000, 10000 and
15000 MCS) of the Monte Carlo time $t$.
(b) Results for $\Delta f_c$ as a function of $t$ at four 
different densities for a $L = 15$ minimum. Results for
other times and densities studied interpolate smoothly with the
results shown. } 
\label{fig1}
\end{figure}

The procedure for determining $\Delta f_c$ is illustrated
in Fig.\ \ref{fig1}a, where we have 
shown the results for the 
transition probability $P$ as a function of the free energy increment
$\Delta f$ for a  
$L = 15$ minimum at $n^* = 0.99$. 
It is clear from our data 
that the uncertainty in the estimated values of $\Delta
f_c$ is $\sim 0.05$, the spacing between successive values of 
$\Delta f$ in the simulation. Typical results for $\Delta f_c$
are shown in Fig.\ \ref{fig1}b for the same $L = 15$ minimum and four 
values of $n^*$. Clearly, $\Delta f_c$ is a
weak function of $t$, and a stronger function of $n^*$.
The dependence of $\Delta f_c$ on $t$ for fixed  $n^*$
becomes more pronounced as $n^*$ is increased. These dependences are
analyzed later in this section. 

The large number of minimization runs we carried out locate a 
large fraction of the full collection of
glassy minima of the free energy. 
For the ``incommensurate'' $L = 15$ sample used in our work, the
number of minima we
have located at each density is in the range of four to six.
The ``commensurate'' $L = 12$ sample exhibits a substantially larger
number of minima, one of which is crystalline (fcc). 
A similar sensitivity of the number of local minima to the sample size and
boundary conditions has been found in numerical studies~\cite{ah97,dpvr} of 
the potential energy landscape of model liquids described 
by simple Hamiltonians. In the following discussion of the statistical
properties of the collection of glassy minima, we consider chiefly the 
results obtained for $L = 12$, for which we can produce significant 
statistics.

\begin{figure}[htbp]
  \begin{center}
    \leavevmode 
    \hspace*{1.8cm}
    \epsfysize=5.5cm\epsfbox{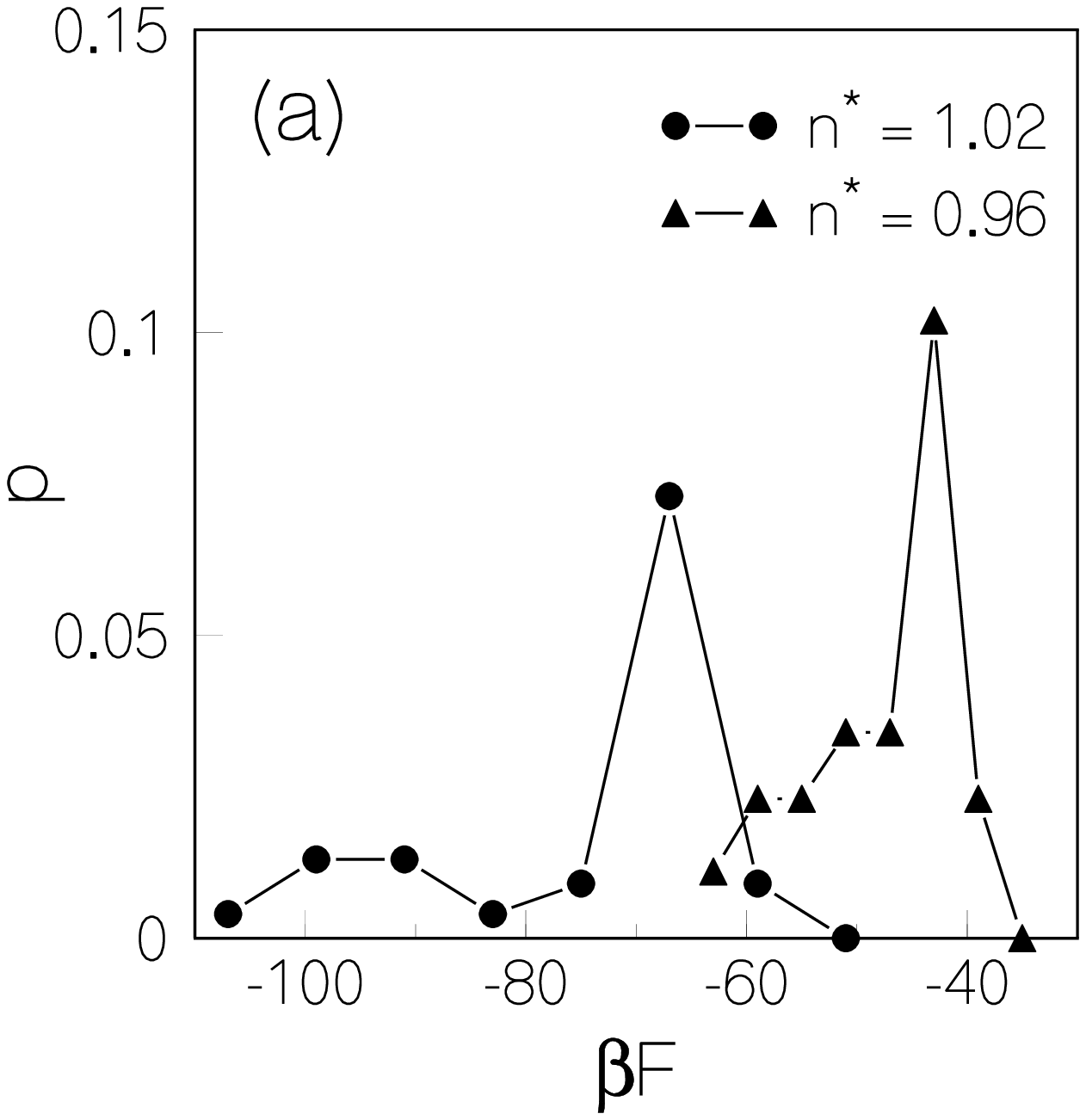}
    \epsfysize=5.5cm\epsfbox{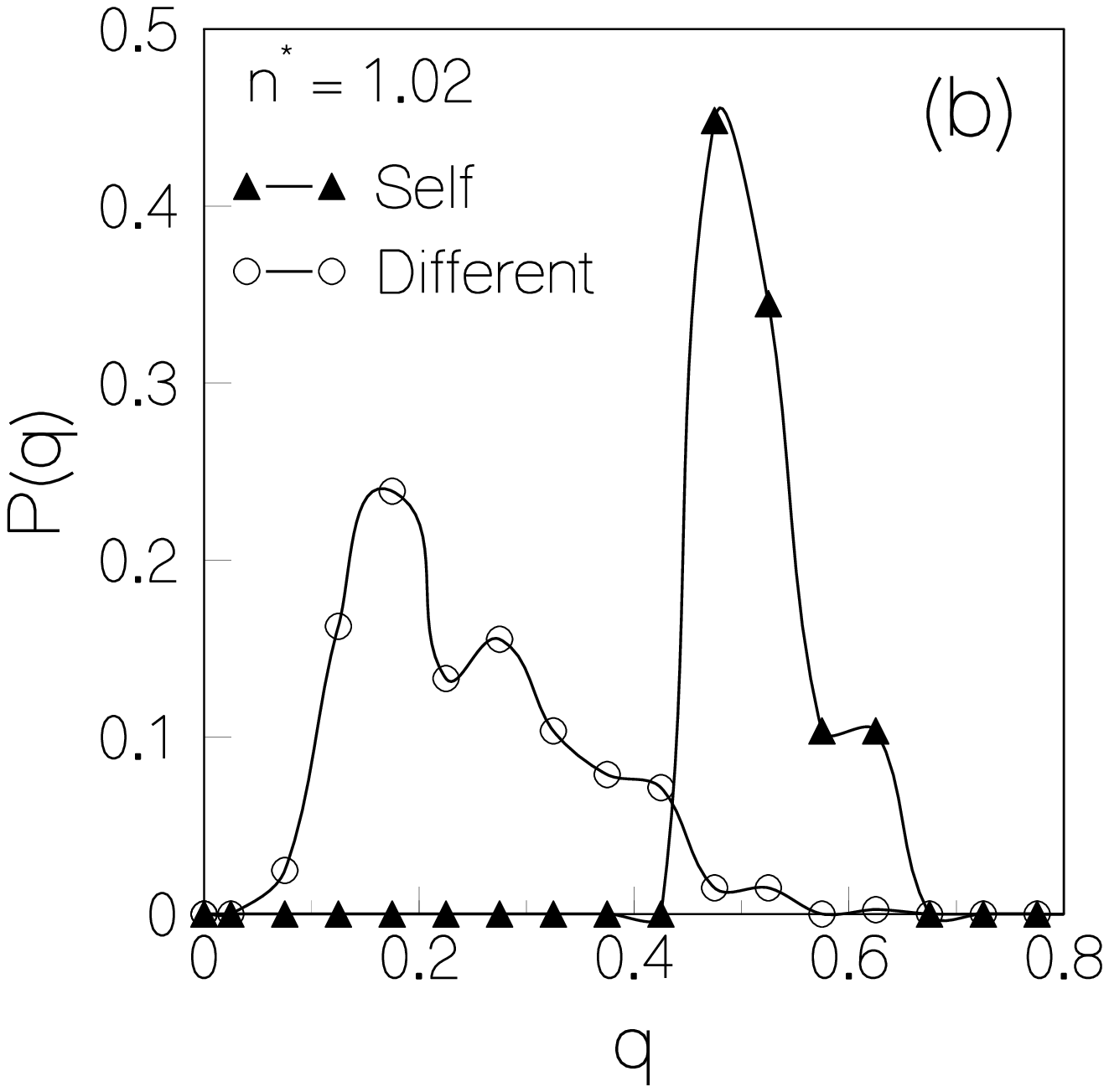}
  \end{center}
  \caption{(a) The ``density of states'' for glassy free energy minima, defined
as the probability of finding a glassy minimum with free energy in
a given range (see text). Results for $L = 12$ samples at two densities
are shown.
(b) The distribution $P(q)$ of the overlap $q$ defined in
Eq.(\protect\ref{ovlp}). The distributions of the self-overlap and the
overlap between different minima are shown separately for $L=12$
samples at density $n^*=1.02$.}
\label{fig2}
\end{figure}

The number of glassy local minima of the $L = 12$ system remains nearly
constant as the density is varied in the range $0.96 \le n^* \le 1.06$.
This number is close to 25. There is no systematic trend in the
dependence of this number on the density. 
The free energies of these minima are distributed in a band that lies
between the free energy of the uniform liquid (zero)
and that of the crystal. The
width of this band increases with $n^*$. Since the number of
minima is approximately independent of the density, this implies that
the ``density of states'' of the glassy minima decreases as $n^*$ is
increased. Let $p(\beta F) \delta $ be the probability of
finding a glassy minimum with dimensionless free energy between $\beta
F-\delta/2$ and $\beta F + \delta/2$. 
We have calculated this quantity 
at different values of $n^*$. Representative results at two
densities, $n^*=0.96$ and $n^* = 1.02$, are shown in Fig.\ \ref{fig2}a.
The values of $\delta$ used are 4.0 and 8.0 for $n^*=0.96$ and $n^* =
1.02$, respectively. The
range of $\beta F$ over which $p(\beta F)$ is nonzero is clearly wider
at the higher density. The consequent decrease in the values of
$p(\beta F)$ with increasing density is also clearly seen. Both
distributions show peaks near the upper end, and tails extending to
substantially lower values. However, the lowest free energy of the
glassy minima is substantially higher than the free energy of the
crystalline minimum. If the probability of
finding the system in a glassy minimum is assumed to be proportional to
the Boltzmann factor $e^{-\beta F}$, then only those minima with free
energies lying near the lower end of the band would be relevant in
determining the equilibrium and dynamic properties of the system. Our
results indicate that the number of such ``relevant'' minima decreases
with increasing $n^*$. We find correlations between the free
energy of a glassy minimum and its structure, similar to those
found in Ref.\cite{dv94}. Minima with
lower free energies have more ``structure'' (as indicated by e.g. the
heights of the first and second peaks of the two-point correlation 
function of the local density) and higher average density than those with
higher free energies.

We have also studied how the distributions of the local density 
variables in two distinct glassy minima differ from one another. 
The degree of similarity between two minima may be quantified in terms
of their ``overlap''~\cite{ktw}. For the discretized system
considered here, the overlap $q(1,2)$ between two minima labeled ``1''
and ``2'' may be defined in the following way:
\begin{equation}
q(1,2) = \frac{1}{\rho_{av}L^3}\,max\{{\bf R}\}
\sum_i[\rho_i^{(1)} - \rho_{av}][\rho_{{\bf R}(i)}^{(2)}-\rho_{av}].
\label{ovlp}
\end{equation}
Here, $\rho_i^{(1)}$ and $\rho_i^{(2)}$ are the discretized densities
at the two minima and $\rho_{av}$ is the average value of the $\rho_i$, which
is assumed to be the same for the two minima.
$\bf R$ represents one of the 48 symmetry operations
of the cubic mesh, plus all translations taking into account periodic
boundary conditions,
${\bf R}(i)$ is the mesh point to which mesh point $i$
is transformed under $\bf R$, and $max\{{\bf R}\}$ means that the $\bf
R$ that maximizes the quantity on the right is to be taken.
In Fig.\ \ref{fig2}b, we display
the results for the distribution $P(q)$ of
$q$ at $n^*=1.02$. The distribution of the self-overlap, defined as
$q(i,i)$ for the $i$th minimum, is also shown. As expected, the
distribution of the self-overlap exhibits a sharp peak at a large
value of $q$. The overlap between different minima exhibits a broad
distribution with a peak at a small value of $q$, indicating that
most of the glassy minima are rather different
from one another. This distribution, however, extends to values of $q$ as
large as 0.6, indicating that there are a few pairs of 
glassy minima which are very similar to each other. For each value of 
$n^*$, we find a small number (3-5) of such pairs of minima.
The main difference between their structures comes from small
displacements of just 2-3 particles. 
These pairs of minima are examples of ``two-level systems'' whose existence
in glassy materials was postulated~\cite{tls} many years ago. 

We have also looked at how the quantity $ F_c = F_0 +\Delta F_c$,
varies from one minimum to another. While the free energy $F_0$ of a glassy
minimum varies over a wide range (see Fig.~\ref{fig2}a), the value of $F_c$
is nearly constant for each value of $n^*$.  This suggests a 
``putting green like'' free energy landscape in which the local minima
are like ``holes'' of varying depth in a nearly flat background. This
structure also implies that there is a strong correlation between the
depth of a minimum and the height of the barriers that separate it from
the other minima: the barriers are higher for deeper minima. 

We now discuss the dependence of $\Delta f_c$ on the density $n^*$ and
MC time $t$. Since the transition probability $P$ is an 
increasing function of both $\Delta f$ and $t$,
$\Delta f_c(n^*,t)$  decreases as $t$ is increased (see Fig.~\ref{fig1}). 
In agreement with the previously observed~\cite{dv96}
growth of the barrier-crossing time scale with $n^*$, we find that 
$\Delta f_c$ is an increasing function of $n^*$. 
The $t$-dependence of $\Delta f_c$ becomes {\em stronger} as $n^*$
is increased. The $t$-dependence of $\Delta f_c$ is
closely related to the probability of finding a path 
(``saddle point'') that connects
the starting minimum to a different one. 
If such paths were relatively easy to find, then the transition 
probability would be insensitive to the value of $t$ as long as
it is not very short. If, however, paths to other minima are few,
a large number of configurations have to be explored 
before one of them is found. The $t$-dependence 
of $\Delta f_c$ would then be more pronounced
and extend to larger values of $t$. To make the idea more concrete, 
we ignore the short-range time correlations  among 
the configurations generated in a
MC run and assume that they represent $t$ independent 
samplings of configurations with
free energy less than $F_0+\Delta F$.  
Neglecting the rare return to the basin of attraction of 
the starting minimum after 
a transition to a different basin of attraction,
the transition probability may then be estimated
as $P(n^*,\Delta f,t) = 1- [1-p(n^*,\Delta f)]^{t} \simeq 1 -
\exp(-tp)$, where 
$p(n^*, \Delta f) \ll 1$ is the probability that a randomly 
chosen configuration with
$\beta F \le \beta F_0 + N \Delta f$ belongs in the basin of attraction of
a different minimum. One expects $p$ to be zero if $\Delta f \le
\Delta f_0(n^*)$ where $N k_B T \Delta f_0$ is the height of the lowest 
free energy barrier, and
$p = g(n^*,\Delta f - \Delta f_0)$ for $\Delta f > \Delta f_0$ where 
$g(n^*,x)$ grows 
continuously from zero as $x$ is increased from zero.
Combining this with the definition of $\Delta f_c$, we obtain the relation 
$g(n^*,\Delta f_c(n^*,t)-\Delta f_0(n^*)) = \ln 2/t$. 
Since $\Delta f_c(n^*,t_1) - \Delta f_c(n^*, t_2)$ for fixed $t_1 < t_2$ 
{\em increases} with $n^*$,  
the function $g(n^*,x)$
{\em decreases} (i.e. the difficulty of finding 
paths to other minima increases) as
$n^*$ is increased at fixed  $x$.

\begin{figure}[htbp]
  \begin{center}
    \leavevmode 
    \hspace*{1.8cm}
    \epsfysize=5.5cm\epsfbox{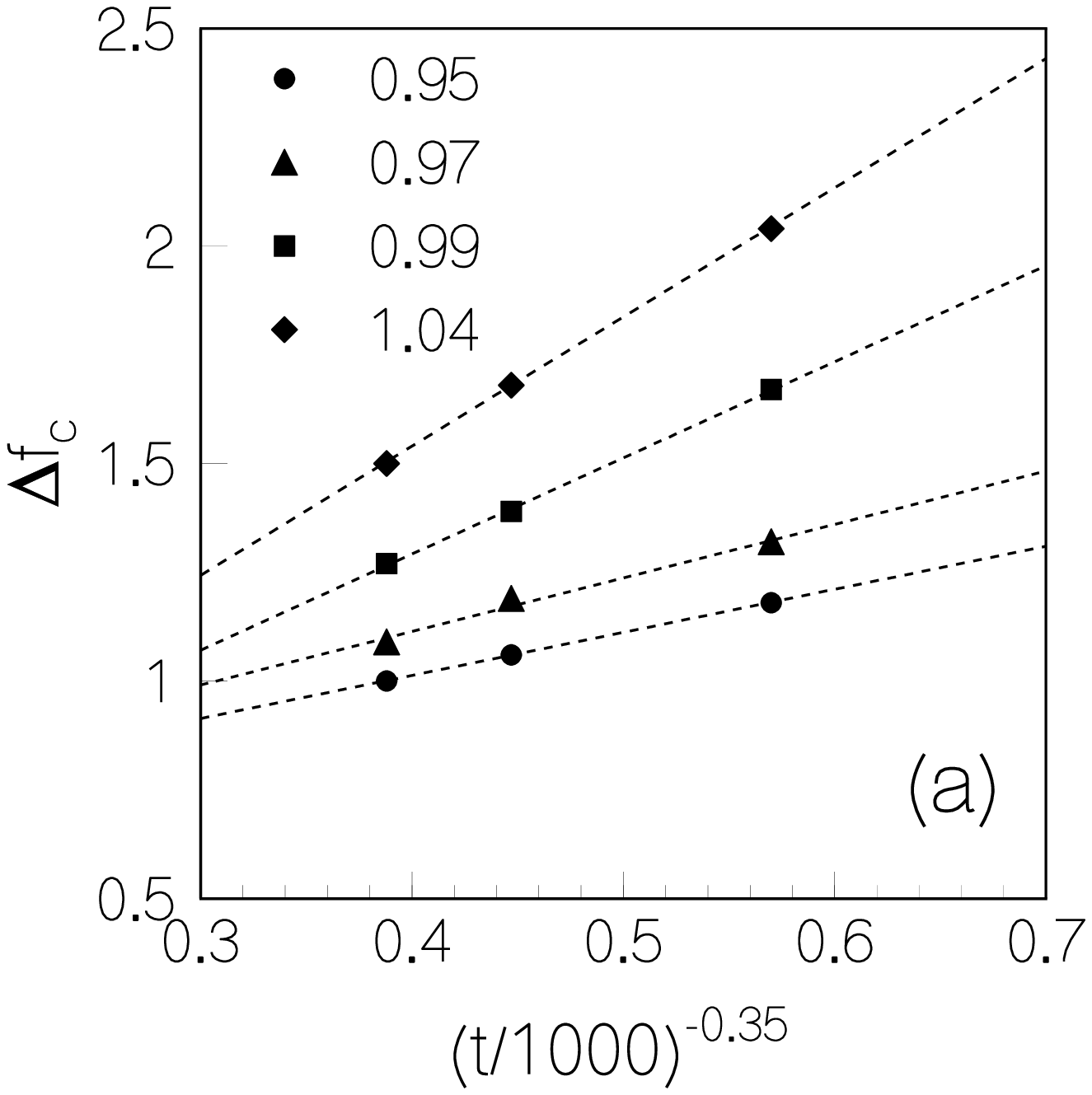}
    \epsfysize=5.5cm\epsfbox{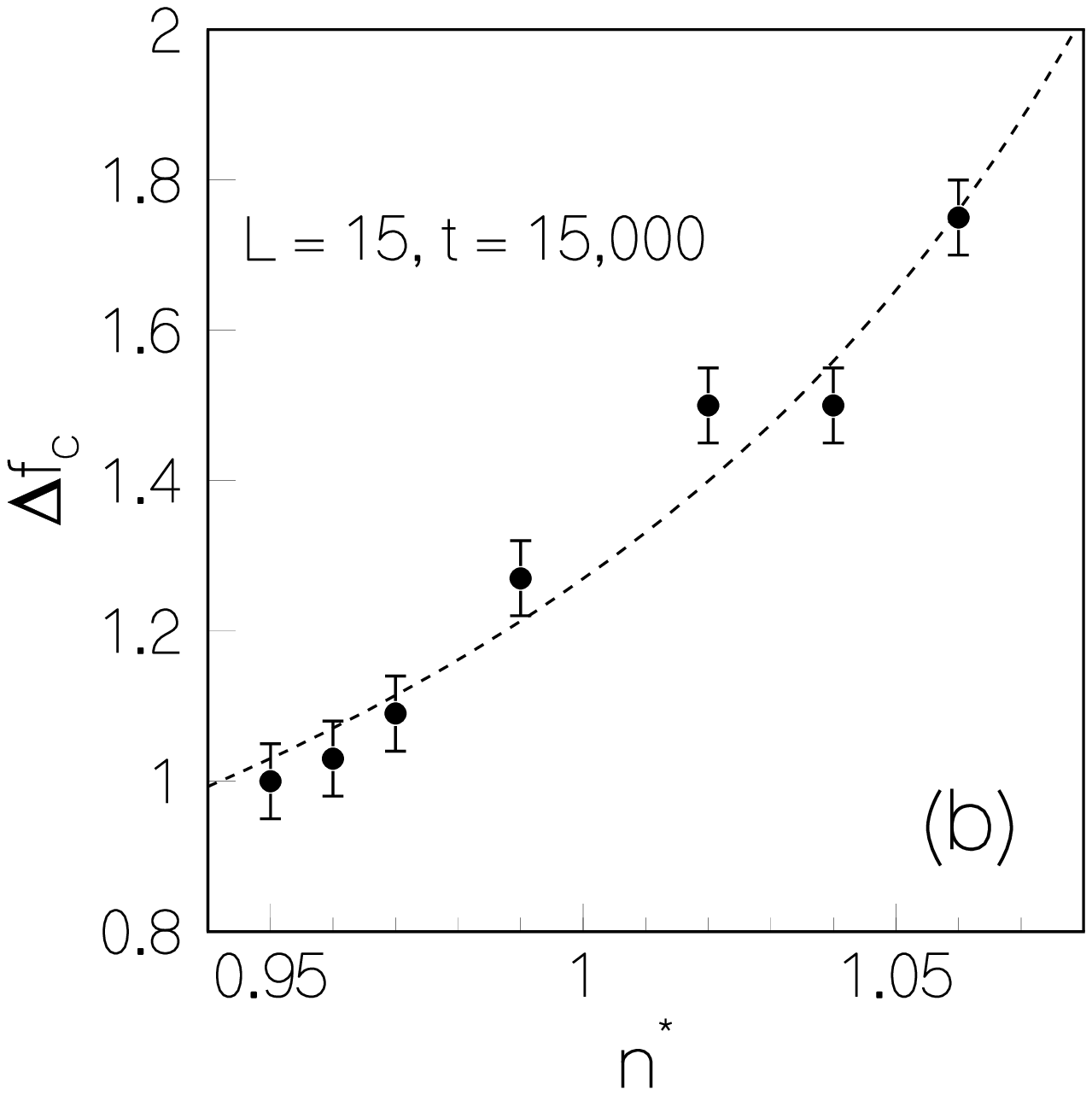}
  \end{center}
  \caption{(a) Plots of $\Delta f_c$, obtained for a $L$ = 15 minimum, against 
$(t/1000)^{-0.35}$ for four values of $n^*$. The dashed lines are the
best straight-line fits.
(b) The dependence of $\Delta f_c(n^*, t=15000)$ on $n^*$ for $L$ = 15. 
The dashed line shows the best fit to a Vogel-Fulcher form (see text).} 
\label{fig3}
\end{figure}

The observed $t$-dependence of $\Delta f_c$ for 
all values of $n^*$
and all the minima in our study is well-represented by 
\begin{equation}
\Delta f_c(n^*,t) = \Delta f_0(n^*) + c(n^*) t^{-\alpha},
\label{fit}
\end{equation}
with $\alpha$ in the range $0.25-0.40$. 
Fits to this form with $\alpha = 0.35$
for a minimum with $L$ = 15 are shown in Fig.~\ref{fig3}a. 
The values of $\Delta f_0$ obtained from such
fits with a fixed value of $\alpha$ are nearly independent of $n^*$, 
but exhibit a
dependence on the value of $\alpha$, varying between 0 and 0.5
for the $L$ = 15 minimum. Similar results are obtained for $L=12$, with
values of $\Delta f_0$ between 1.3 and 1.5.
The quantity $c(n^*)$ increases with $n^*$. 
These results correspond to 
$g(n^*,x) \sim A(n^*)x^{1/\alpha}$ with $A(n^*)$ 
decreasing with increasing $n^*$. 
We conclude that the 
growth of the effective barrier height
with increasing $n^*$ is primarily due to an 
entropic mechanism associated with an increase
of the difficulty in finding low-lying saddle points that connect different 
glassy local minimum of the free energy. 

The dependence of $\Delta f_c$ on $n^*$ 
is consistent with the Vogel-Fulcher
law~\cite{vf} which assumes the following form~\cite{wa81} for our system:
\begin{equation}
\Delta f_c(n^*) = a + b/(n^*_c - n^*),
\label{vfeq}
\end{equation}
where $a$, $b$ and $n^*_c$ are constants. 
The value of $n^*_c$ obtained
from fits of our data for $\Delta f_c(n^*, t)$ to Eq.(\ref{vfeq}) 
with fixed $a$ is
nearly independent of $t$. This is
consistent with the form of Eq.(\ref{fit}) if 
$a = \Delta f_0$ , $b \propto t^{-\alpha}$, and $c \propto 1/(n^*_c-n^*)$. 
$\Delta f_0$ is indeed nearly independent of
$n^*$, and we find that
the $t$-dependence of $b$ and the $n^*$-dependence of $c$ 
are in agreement
with the other two conditions.  For the $L$ = 15 case,
we can fit the data for $\Delta f_c$ at $t$ = 15,000 to 
the form of Eq.(\ref{vfeq}) with
$a$ = 0 ($\Delta f_0=0)$. 
The best fit, shown in Fig.~\ref{fig3}b, 
corresponds to $n^*_c = 1.225$,  very
close to the expected random close packing density 
$n^*_{rcp} \simeq 1.23$. 
The best fit to the $L$ = 12 data with $a \simeq 1.0$ 
also yields a similar value of $n^*_c$. 
We conclude that the observed growth of
the effective barrier height is consistent with the Vogel-Fulcher form. 
The increase in the effective barrier height as $n^*$ is increased from
0.96 to 1.06 is about 25$k_BT$, corresponding to a growth of the 
characteristic time
scale of about ten orders of magnitude. Thus, the range of time 
scales covered in
our study is comparable to that used in Vogel-Fulcher fits 
of experimental data, and
much wider than what can be achieved in standard MC or 
molecular dynamics simulations.

\section{Discussion}
\label{s&d}

We close with a discussion of the connections between our
results and those of 
spin-glass-like theories~\cite{kt,ktw,gp1} of the structural 
glass transition (see Ref.~\cite{dv99} for a discussion of the relation
of our work with other recent studies of the behavior of simple liquids
near the glass transition). These theories are based on the similarity 
between the phenomenology of the structural glass transition in 
so-called ``fragile''~\cite{rev2} liquids and the behavior found in a class
of generalized mean-field spin glass models~\cite{kw,ktn} with 
infinite-range interactions.
At high temperatures, the free energy of these mean-field models,
expressed as a function of the single-site magnetizations, 
exhibits only one, ``paramagnetic'', minimum. As
$T$ is lowered, an exponentially large number of
non-trivial local minima come into existence at a
temperature $T_d$, where a ``dynamic transition'', 
characterized by a breaking of ergodicity, occurs. 
This ``dynamic transition'' does not have any signature in
the equilibrium behavior of the system. A thermodynamic phase
transition occurs at a lower temperature $T_c$. 
In the suggested analogy between these models and the structural glass
transition, the paramagnetic minimum of the free energy is identified
with that corresponding to the uniform liquid, and the role of the 
non-trivial local minima of the free energy is played by the glassy
local minima. The analogue of the ``dynamic transition'' at $T_d$ 
is thought to be smeared out in liquids.  It has been 
suggested~\cite{kt,ktw,gp1} that  $T_d$ should be identified with the
``ideal glass transition'' temperature of mode-coupling
theories~\cite{mct}. 
The temperature $T_c$ is interpreted as the ``Kauzmann 
temperature''~\cite{kauz} at which the difference in entropy 
between the supercooled 
liquid and the crystalline solid extrapolates to zero. The relaxation
time of the supercooled liquid is supposed to diverge at this
temperature. Heuristic arguments  suggest that this divergence is
of the Vogel-Fulcher form~\cite{ktw,gp1}.

Our results qualitatively support this scenario. We find a
characteristic density at which a large number of glassy
minima of the free energy appear. We do not 
know whether the number of glassy minima depends exponentially on the
sample volume. The configurational entropy associated with these
minima decreases with increasing density because the width of the band
over which the free energy of these minima is distributed increases
with density. We have also found evidence for a
Vogel-Fulcher-type growth of relaxation times driven by an entropic 
mechanism. 

There are, however, certain differences between our
findings and the predictions of spin-glass-like theories. In our
earlier work \cite{dv94,vd95}, we found that the free energy of a typical
glassy minimum becomes lower than that of the uniform liquid as the
density is increased slightly above the value ($n^* \approx$ 0.8) at 
which the minimum comes into existence. In particular, the free
energies of the glassy minima are substantially lower than that of the
uniform liquid one for $n^*$ near $n^*_x \simeq 0.95$, the crossover 
density~\cite{dv94,vd95} above which the dynamics is governed by 
transitions among glassy minima. This is different from the
behavior found in the spin glass models. Our results for the
distribution of the overlap between different minima are also somewhat
different from those for the spin glass models. Some of these
differences may be due to finite-size effects which are probably 
significant
for the small samples considered here. Also, fluctuation effects, which
are unimportant in mean-field models, may play an important role in our
system. A careful investigation of these issues would be interesting.


\section*{References}
\begin{thebibliography}{999}
\bibitem{rev1} J\"{a}ckle J 1986 {\it Rep. Prog. Phys.} {\bf 49} 171.
\bibitem{rev2} Angell C A 1988 {\it J. Phys. Chem. Solids} {\bf 49} 863.
\bibitem{pwa} Anderson P W 1979 in {\it Ill Condensed Matter, Lecture
Notes of the Les Houches Summer School}, ed R Balian, R Maynard and G 
Toulouse (Amsterdam: North Holland). 
\bibitem{pgw} Wolynes P G 1988 in {\it Proceedings of International
Symposium on Frontiers in Science, (AIP Conf. Proc. No. 180)}, ed
S S Chen and P G Debrunner (New York: American Institute of Physics).
\bibitem{kw} Kirkpatrick T R and  Wolynes P G 1987 {\it Phys. Rev. A}
{\bf 35} 3072; {\it Phys. Rev B} {\bf 36} 8552.
\bibitem{kt} Kirkpatrick T R and Thirumalai D 1989 {\it J. Phys. A}
{\bf 22} L149.
\bibitem{ktw} Kirkpatrick T R, Thirumalai D and Wolynes P G 1989
{\it Phys. Rev. A} {\bf 40} 1045.
\bibitem{bm} Bouchaud J -P and Mezard M 1994 {\it J. Phys. I (France)}
{\bf 4} 1109.
\bibitem{parisi} Cugliandolo L F, Kurchan J, Parisi G and 
Ritort F 1995 {\it Phys. Rev. Lett.} {\bf 74} 1012.
\bibitem{RY} Ramakrishnan T V and Yussouff M  1979 {\it Phys. Rev. B}
{\bf 19} 2775.
\bibitem{cd1} Dasgupta C 1992 {\it Europhys. Lett.} {\bf 20} 131.
\bibitem{cd2} Dasgupta C and Ramaswamy S 1992 {\it Physica A} {\bf 186}
314.
\bibitem{lvd} Lust L M, Valls O T, and Dasgupta C 1993 {\it Phys. Rev.
E} {\bf 48} 1787.
\bibitem{dv94} Dasgupta C and Valls O T 1994 {\it Phys Rev. E} {\bf 50}
3916.
\bibitem{vd95} Valls O T and Dasgupta C 1995  {\it Transport Theory
and Stat. Physics} {\bf 24} 1199.
\bibitem{dv96} Dasgupta C and Valls O T 1996 {\it Phys. Rev. E} {\bf
53} 2603.
\bibitem{tls} Anderson P W, Halperin B I and Varma C M 1972 {\it
Philos. Mag.} {\bf 25} 1; Phillips W A 1972 
{\it J. Low. Temp. Phys.} {\bf 7} 351.
\bibitem{vf} Vogel H 1921 {\it Z. Phys.} {\bf 22} 645; Fulcher G S
1925 {\it J. Amer. Ceram. Soc.} {\bf 8} 339.
\bibitem{wa81} Woodcock L V and Angell C A 1981 {\it Phys. Rev. Lett.} 
{\bf 47} 1129.
\bibitem{hm86} Hansen J P and McDonald I R 1986 {\it Theory of Simple 
Liquids} (London: Academic).
\bibitem{dv98} Dasgupta C and Valls O T 1998 {\it Phys. Rev. E} {\bf
58} 801.
\bibitem{dv99} Dasgupta C and Valls O T 1998 {\it Phys. Rev. E} {\bf
59} 3123.
\bibitem{ah97} Heuer A 1997 {\it Phys. Rev. Lett.} {\bf 78} 4051.
\bibitem{dpvr} Daldoss G, Pilla O, Villani G, and Ruocco G, 1998
preprint (cond-mat/9804113).
\bibitem{gp1} Parisi G 1997 in {\it Complex Behaviour of Glassy Systems:
Proceedings of the XIV Sitges Conference}, ed M Rubi and C
Perez-Vicente (Berlin: Springer).
\bibitem{ktn} Kirkpatrick T R and Thirumalai D 1987 {\it Phys. Rev. B} 
{\bf 36} 5388.
\bibitem{mct} G\"{o}tze W 1991 in {\it Liquids, Freezing and the Glass
Transition} ed J P Hansen, D Levesque and J Zinn-Justin (New York: Elsevier).
\bibitem{kauz} Kauzmann W 1948 {\it Chem. Rev.} {\bf 48} 219.

\end{thebibliography}
\end{document}